\documentclass[conference,10pt]{IEEEtran}
\IEEEoverridecommandlockouts
\usepackage{cite}
\usepackage{amsmath,amssymb,amsfonts}
\usepackage{algorithmic}
\usepackage{graphicx}
\usepackage{textcomp}
\usepackage{xcolor}
\usepackage{array}
\usepackage{multirow}
\usepackage{enumitem}
\usepackage{makecell}
\usepackage{subcaption}
\usepackage{booktabs}

\usepackage{caption}
\newcolumntype{B}{>{\raggedright\arraybackslash}m{2 cm}}
\newcolumntype{?}{!{\vrule width 1pt}}

\def\BibTeX{{\rm B\kern-.05em{\sc i\kern-.025em b}\kern-.08em
		T\kern-.1667em\lower.7ex\hbox{E}\kern-.125emX}}
\usepackage[letterpaper, portrait, margin=0.7in, bmargin=1in]{geometry}

\usepackage{fancyhdr}
\fancypagestyle{firststyle}{\fancyhf{}
		\fancyhead[L]{\small Dipankar Shakya, Mingjun Ying, Theodore S. Rappaport, Hitesh Poddar, Peijie Ma, Yanbo Wang, and Idris Al-Wazani, ``Wideband Penetration Loss through Building Materials and Partitions at 6.75 GHz in FR1(C) and 16.95 GHz in the FR3 Upper Mid-band spectrum",\textit{ (Submitted) GLOBECOM 2024 - 2024 IEEE Global Communications Conference, }South Africa, Dec. 2024, pp. 1--6.}
	}

\begin{document}
	\bstctlcite{IEEEtran:BSTcontrol}
	\title{%Characterizing Penetration Loss with a Wideband Radio Channel Measurement System in the Upper Mid-band.
		%Measurement-based Characterization of Penetration Loss for Common Indoor Materials and Partitions at 6.75 GHz and 16.95 GHz in the Upper Mid-band
		Wideband Penetration Loss through Building Materials and Partitions at 6.75 GHz in FR1(C) and 16.95 GHz in the FR3 Upper Mid-band spectrum
	}

	\author{\IEEEauthorblockN{Dipankar Shakya$^{\dagger1}$, Mingjun Ying$^{\dagger2}$, Theodore S. Rappaport$^{\dagger3}$, \\Hitesh Poddar*, Peijie Ma$^\dagger$, Yanbo Wang$^\dagger$, and Idris Al-Wazani$^\dagger$}
		\IEEEauthorblockA{\textit{$^\dagger$NYU WIRELESS, Tandon School of Engineering, New York University, USA}\\
			\IEEEauthorblockA{\textit{*Sharp Laboratories of America (SLA), Vancouver, Washington, USA}}
			\{$^1$dshakya, $^2$yingmingjun, $^3$tsr\}@nyu.edu}
		\thanks{This research is supported by the New York University (NYU) WIRELESS Industrial Affiliates Program.}
	}
	
	\maketitle
	
	\linespread{1.05}
	
	\thispagestyle{firststyle}
	
	\begin{abstract}
		The 4--8 GHz FR1(C) and 7--24 GHz upper mid-band FR3 spectrum are promising new 6G spectrum allocations being considered by the International Telecommunications Union (ITU) and major governments around the world. There is an urgent need to understand the propagation behavior and radio coverage, outage, and material penetration for the global mobile wireless industry in both indoor and outdoor environments in these emerging frequency bands. This work presents measurements and models that describe the penetration loss in co-polarized and cross-polarized antenna configurations, exhibited by common materials found inside buildings and on building perimeters, including concrete, low-emissivity glass, wood, doors, drywall, and whiteboard at 6.75 GHz and 16.95 GHz. Measurement results show consistent lower penetration loss at 6.75 GHz compared to 16.95 GHz for all ten materials measured for co and cross-polarized antennas at incidence. For instance, the low-emissivity glass wall presents 33.7 dB loss at 6.75 GHz, while presenting 42.3 dB loss at 16.95 GHz. Penetration loss at these frequencies is contrasted with measurements at sub-6 GHz, mmWave and sub-THz frequencies along with 3GPP material penetration loss models. The results provide critical knowledge for future 5G and 6G cellular system deployments as well as refinements for the 3GPP material penetration models.                
	\end{abstract}

	\begin{IEEEkeywords}
		3GPP, 6G, FR3, FR1(C), penetration loss, XPD, upper mid-band, upper 6 GHz, partition loss, materials 
	\end{IEEEkeywords}
	
	\section{Introduction}
	The desire to achieve the wide bandwidths and massive data throughput of the mmWave and sub-THz frequencies, while maintaining improved coverage with relatively low gain antennas at sub-6 GHz frequencies has invited strong interest in the FR1(C) and FR3 upper mid-band spectrum. Often heralded as the ``golden band" due to the promising balance between expansive coverage using low gain antennas and smaller partition losses, and becoming closer to offering high capacity that occurs at mmWave and sub-THz bands due to wide bandwidths. The 4--8 GHz FR1(C) and 7--24 GHz FR3 upper mid-band spectrum bands are anticipated to play a crucial role in the deployment of next-generation 5G and 6G cellular systems\cite{Shakya2024gc,Kang2024OJCOM}. Agencies such as ITU and the National Telecommunications and Information Administration (NTIA) are rigorously analyzing these bands to ensure effective utilization alongside current incumbents such as satellite communications, earth exploration, and radio astronomy \cite{NTIA2024}. Endeavors by the 3$^{rd}$ Generation Partnership Project (3GPP) and the Federal Communication Commission (FCC) towards exploration of the FR3 frequencies underscore their critical role in meeting the escalating data requirements and setting the stage for future cellular standards \cite{3GPP3820}. However, there is limited knowledge regarding the radio propagation characteristics at these frequencies for cellular deployments. Particularly, penetration through various building materials and partitions, remain under-explored.
	
	%Particularly, ITU and NTIA have put specific focus on the 7.125-8.4 GHz band, and other segments including the 4.40-4.80 GHz and 14.8-15.35 GHz, as identified by the ITU World Radio Conference 2023 (WRC-23) \cite{NTIA2024}. Furthermore, the commercial significance of this spectrum is acknowledged by industry leaders, such as Nokia \cite{Nokia6GSpectrum}, Huawei \cite{Huawei2021}, and the Alliance for Telecommunications Industry Solutions (ATIS), who have identified crucial sub-bands for future network developments. Endeavors by the 3$^{rd}$ Generation Partnership Project (3GPP) and the Federal Communication Commission (FCC) towards exploration of the FR3 frequencies underscore their critical role in meeting the escalating data requirements and setting the stage for future cellular standards \cite{3GPP3820}. However, there is limited knowledge regarding the propagation characteristics of electromagnetic waves at these frequencies for cellular deployments. Particularly, penetration through various building materials and partitions, remain under-explored. 
	
	\par Extensive research efforts have endeavored to characterize the penetration loss of various materials in mmWave \cite{rodriguez:2017:an-empirical-Outdoor-to-Indoor, diakhate:2017:millimeter-wave-outdoor-to-indoor} and sub-6 GHz bands \cite{durgin:1998:measurements-and-models-for-radio, schwengler:2000:propagation-models-at-5.8-GHz-path-loss}. However, investigation into the penetration loss of various materials found in indoor and outdoor scenarios for the upper mid-bands have been limited. Muqaibel \textit{et al.} \cite{muqaibel:2005:ultrawideband-through-the-wall} studied the
	propagation of ultrawideband (UWB) signals through common building materials such as drywall, plywood, wooden door, glass, brick wall, concrete block wall, styrofoam, office cloth partition and reinforced concrete wall at frequencies between 2-10 GHz. Authors in \cite{r12403280} measured penetration loss of wood panels 0.6, 1, and 1.4 cm thick using a narrowband swept network analyzer from 7--15 GHz and reported 1.9, 1.7, and 3.2 dB loss at 7 GHz and 2, 4.1, and 5 dB loss at 15.5 GHz. Authors in \cite{muqaibel:2005:ultrawideband-through-the-wall} measured the dielectric properties of materials (permittivity and loss tangent) to characterize attenuation and distortion of UWB signals when propagating through the stated materials. Landron \textit{et al.}\cite{Landron1996tap} measured reflection coefficients of limestone, glass, and brick wall surfaces at 1.9 GHz and 4 GHz for vertical and horizontal antenna polarizations, and found Freshnel reflection coefficient models with rough surface correction factors are adequate to model reflection.  Rodriguez \textit{et al.} \cite{rodriguez:2014:radio-propagation-into-modern-buildings} conducted measurements to study the outdoor-to-indoor (O2I) attenuation for buildings in Aalborg, Denmark using a signal generator to sweep a single tone from 800 MHz to 18 GHz, and found modern buildings presented increased penetration loss of 20-25 dB compared to old constructions. Zhang \textit{et al.} \cite{zhang:1994:measurements-of-the-characteristics-of} characterized the penetration loss from 0.9--18 GHz for indoor partitions such as reinforced concrete and plasterboard walls. Measured results in \cite{zhang:1994:measurements-of-the-characteristics-of} showed increased penetration loss with frequency for reinforced concrete wall, however, a monotonic increase in penetration loss with frequency was not observed for plasterboard walls. %Ökvist \textit{et al.} \cite{okvist:2015:15-GHz-propagation-properties} presented penetration loss measurements in an urban environment at 15 GHz and reported losses of 24 dB for triple-pane infrared reflective (IRR) glass. 
	
	In this paper, results of penetration loss measurements for several indoor materials at 6.75 and 16.95 GHz for co and cross-polarized antenna configurations are presented, using a 1 GHz wideband spread spectrum channel sounder (Section \ref{sxn:chS}) \cite{Mac2017jsac,Shakya2021tcas}. The wideband channel sounder avoids frequency selective readings at specific frequencies, which can result in anomalous penetration measurements due to material frequency selectivity or multipath with similar propagation delays in the environment \cite{Newhall1996rd}. Measurements are carried out in the 370 Jay Street building, NYU Tandon School of Engineering, Brooklyn, NY. Calibration procedures and resulting linear operation range of the channel sounder is detailed in Section \ref{sxn:cal}. The antenna cross-polarization discrimination (XPD) for the horn antennas used at 6.75 and 16.95 GHz are characterized in Section \ref{sxn:xpd}. Section \ref{sxn:msmt} highlights the measurement procedure for obtaining the penetration losses. Observations and resulting penetration loss values for co and cross polarized antenna configurations are discussed in Section \ref{sxn:res}. Measurements are compared with 3GPP standard models in Section \ref{sxn:Comp} before the conclusion.

	\section{Wideband Sliding Correlation Channel Sounder for FR1(C) and FR3 Measurements}
	\label{sxn:chS}
\begin{figure}[htbp]
		\centering
		\includegraphics[width=0.5\textwidth]{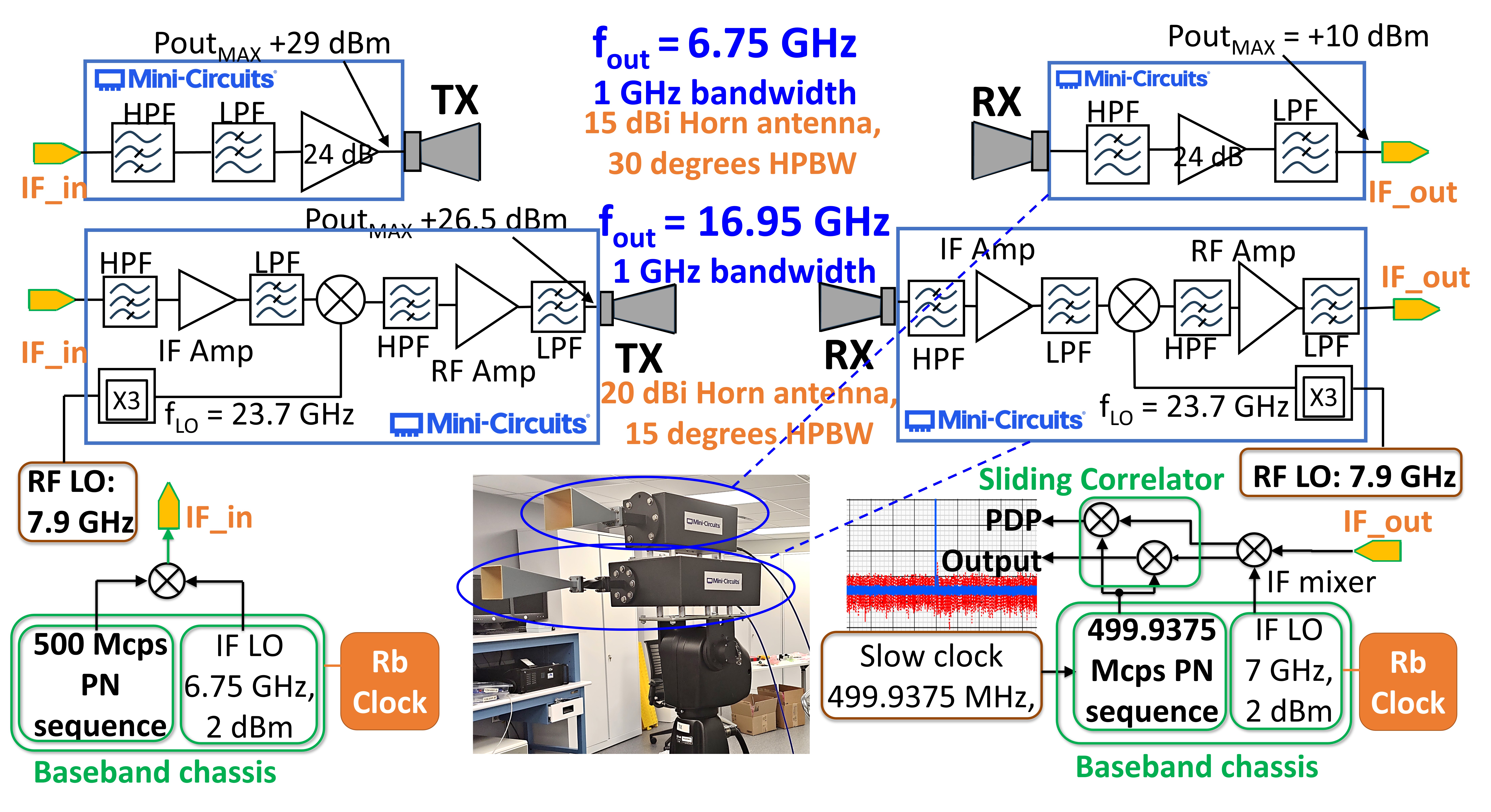}
		\vspace{-15 pt}\caption{The FR1(C) and FR3 wideband channel sounder system with co-located RF front-end modules developed by Mini-Circuits. Top module operates at 6.75 GHz and the bottom module at 16.95 GHz. \newline}
		\label{fig:SysBlock}
		\vspace{-25 pt}
	\end{figure}

	A time-domain channel sounder based on sliding correlation of pseudorandom noise (PN) sequences was employed for accurate mid-band wideband channel propagation and penetration measurements \cite{Shakya2021tcas,Mac2017jsac}.  As presented in Fig. \ref{fig:SysBlock}, the 500 Mcps baseband PN sequence phase modulates a 6.75 GHz carrier to result in a 1 GHz RF bandwidth (BW) signal that is connected to one of two front-end modules, custom developed by Mini-circuits. The distinctive co-located configuration allows a simple transition between the 6.75 GHz and 16.95 GHz frequency bands through the change of a single cable.
	
	The TX front-end module operating at 6.75 GHz amplifies, filters, and transmits the broadband signal through a 15 dBi gain and a 30-degree half-power beamwidth (HPBW) horn antenna, achieving an effective isotropic radiated power (EIRP) of 31 dBm to stay within the licensed emission power level of 35 dBm. The 16.95 GHz TX front-end module employs a heterodyne approach to upconvert the 6.75 GHz intermediate frequency (IF) signal to a RF signal at 16.95 GHz with 1 GHz bandwidth\cite{Shakya2024gc}. The wide bandwidth of 1 GHz allows the channel sounder to overcome frequency selectivity that narrowband and swept narrowband systems are prone to and resolve close multipath in temporal domain to ensure accurate penetration measurements. The polarization of the horn antennas are changeable between vertical (V) and horizontal (H) polarization using waveguide twists at both bands. A total of four antenna configurations V-V or H-H co-polarized, and V-H or H-V cross-polarized are used for measurement of each material under test (MUT). 
	
	%The TX and RX front-end modules use a 7.9 GHz local oscillator (LO) signal, fed into a tripler, yielding a 23.7 GHz output that is subsequently fed into a high-side injection mixer that produces a 16.95 GHz signal. The signal, after upconversion, is band-pass filtered through a high-pass and low-pass filter cascade, and transmitted using a horn antenna with a gain of 20 dBi, maintaining an EIRP of 31 dBm. At the receiver, the signal, once down-converted, is passed through a sliding correlator with a 499.9375 Mcps PN sequence that results in a time-dilated power-delay profile (PDP). The specifications of the channel sounding system are detailed in Table \ref{tab:SysParams}.
	
	\section{Calibration of the Channel Sounder System for Penetration Loss Measurements}
	\label{sxn:cal}
	Calibration ensures the accuracy of the channel impulse response measurements and path loss results, and are conducted at the beginning and end of each measurement day to ensure power levels obtained during the penetration tests are consistent throughout the day. The calibration determines the system gain and linear range of received powers through identification of the correct attenuator settings at the TX and RX, and power levels of the LO signals. Particularly, the attenuator settings determined from the calibration are adjusted during penetration measurements to ensure the received power falls in the linear range of operation, especially when measuring the cross-polarized antenna configurations.   
	
	The correct settings are ascertained through a free-space path loss (FSPL) measurement at a fixed transmitter-receiver (T-R) separation distance, exceeding 5 $\times$ far-field distance ($D_f$) of the horn antenna (here, 4 m T-R separation is used), and antenna height (here, 1.5 m height is used), avoiding ground or ceiling reflected multipath, adhering to the criteria described in \cite{Xing2018vtc}. The transmit power before the TX horn antenna is captured with a power-meter prior to initiating the calibration process. %Fig. \ref{fig:Cal} shows the calibration results of the 16.95 GHz front-end. 
	During calibration, power under the first PDP peak yields the total received power in the LOS multipath for different configurations of the TX and RX attenuator and LO input powers \cite{Xing2018vtc,newhall1997amtams}. Varying the TX and RX attenuators allows for a total 50 dB linear operation range during the penetration measurements over a 150 dB power range. Calibration at the end of day typically shows differences of under 0.2 dB for FSPL.

	\renewcommand{\arraystretch}{1.2}
	\begin{table}[htbp]
		\centering
		\caption{System Parameters for the Upper-Mid Band Channel Sounder at NYU WIRELESS}
		\vspace{-1 em}
		\begin{tabular}{|p{3.2 cm}|>{\centering\arraybackslash}p{2 cm}|>{\centering\arraybackslash}p{2 cm}|}
			\hline
			\raggedleft{\textbf{Carrier Frequency }} & \textbf{6.75 GHz} & \textbf{16.95 GHz} \\
			\hline
			\raggedleft\textbf{Free Space PL at 1m reference distance} & \vfil 49 dB & \vfil 57 dB \\
			\hline
			\raggedleft \textbf{Baseband signal} & \multicolumn{2}{c|}{11th order PN sequence (2047 chips)} \\
			\hline
			\raggedleft \textbf{TX PN Code Chip Rate} & \multicolumn{2}{c|}{500 Mcps} \\
			\hline
			\raggedleft \textbf{TX PN Code Chip Width} & \multicolumn{2}{c|}{2.0 ns} \\
			\hline
			\raggedleft \textbf{RX PN Code Chip Rate} & \multicolumn{2}{c|}{499.9375 Mcps} \\
			\hline
			\raggedleft \textbf{Slide factor} & \multicolumn{2}{c|}{8000} \\
			\hline
			\raggedleft \textbf{Digitizer Sampling rate at RX Oscilloscope} & \multicolumn{2}{c|}{2.5 Msps} \\
			\hline
			\raggedleft \textbf{RF BW (Null-to-null)} & \multicolumn{2}{c|}{1 GHz} \\
			\hline
			\raggedleft \textbf{Max Transmit Power (fed into the horn antenna)} & 29 dBm & 26.5 dBm \\
			\hline
			\raggedleft \textbf{TX/RX Antenna Type} & \multicolumn{2}{c|}{Pyramidal Horn Antenna} \\
			\hline
			\raggedleft \textbf{TX/RX Antenna Dim.} & 3.75"$\times$2.65" &3.08"$\times$2.33" \\
			\hline
			\raggedleft \textbf{TX/RX Antenna Far-field} & 41 cm & 69 cm \\
			\hline
			\raggedleft \textbf{TX/RX Antenna Gain} & 15 dBi & 20 dBi \\
			\hline
			\raggedleft \textbf{TX/RX Ant. HPBW (Az/El)} & \vfil 30$^{\circ}$ / 30$^{\circ}$ & \vfil 15$^{\circ}$ / 15$^{\circ}$ \\
			%investigate HPBW for 73 GHz
			\hline
			\raggedleft\textbf{XPD} & 35 dB & 38 dB \\
			\hline
			\raggedleft\textbf{Max EIRP} & 44 dBm & 46.5 dBm \\
			\hline
			\raggedleft\textbf{Max EIRP used} & \multicolumn{2}{c|}{31 dBm} \\
			\hline
			\raggedleft \textbf{Max Measurable Path Loss (at 5 dB SNR)} & \vfil 155.6 dB & \vfil 159.2 dB \\
			\hline
			\raggedleft \textbf{TX Polarization} & \multicolumn{2}{c|}{Vertical/Horizontal} \\
			\hline
			\raggedleft \textbf{RX Polarization} & \multicolumn{2}{c|}{Vertical/Horizontal} \\
			\hline
			\raggedleft \textbf{TX/RX Waveguide Size} & WR137 & WR62\\
			\hline
		\end{tabular}%
		\label{tab:SysParams}%
		\vspace*{-1\baselineskip}
	\end{table}
	\renewcommand{\arraystretch}{1}
	
	%	\begin{figure}[htbp]
		%		\centering
		%		\includegraphics[width=0.44\textwidth]{fig/Cal.jpg}
		%		%\vspace{-20 pt}
		%		\caption{Results of the channel sounder calibration for 16.95 GHz operation. The system has a linear operating range of 50 dB.\newline}
		%		\label{fig:Cal}
		%		\vspace{-20 pt}
		%	\end{figure}
	
	%	Time calibration helps achieve absolute timing using a time-synchronization algorithm based on the precision-time protocol (PTP). Before measurements, the Rubidium (Rb) clocks on the TX and RX are cable-connected for over 12 hours to correct phase and frequency offsets using a 1 pulse-per-second beat signal. During measurements, as the absolute timing accuracy of the system drifts---a phenomenon termed as time drift---due to clock errors and system noise, the PTP synchronization algorithm detailed in \cite{Shakya2023gc}, minimizes the time drift by multiple orders of magnitude. Furthermore, a successive drift correction algorithm that successively measures a reference PDP at a TX-RX location to recover absolute timing allows achieving absolute timing of multipath during propagation measurements\cite{Shakya2023gc}.
	
	\section{Antenna Cross-Polarization Discrimination}
	\label{sxn:xpd}
\begin{figure}[htbp]
		\centering%
		\subfloat[]{%
			\centering
			\includegraphics[width=45mm]{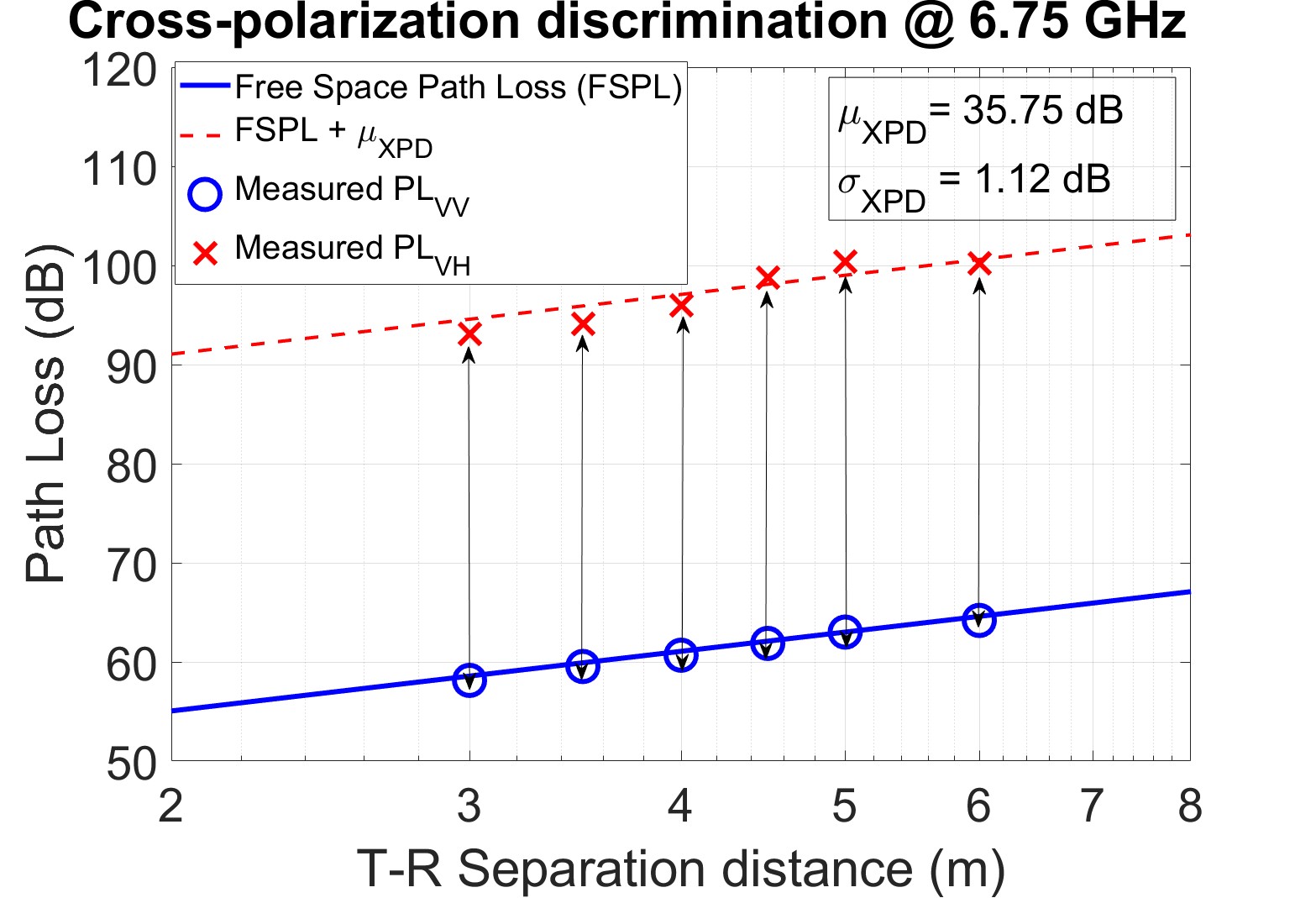}}%
		%\\[-0.1mm]%
		\subfloat[]{%
			\centering
			\includegraphics[width=45mm]{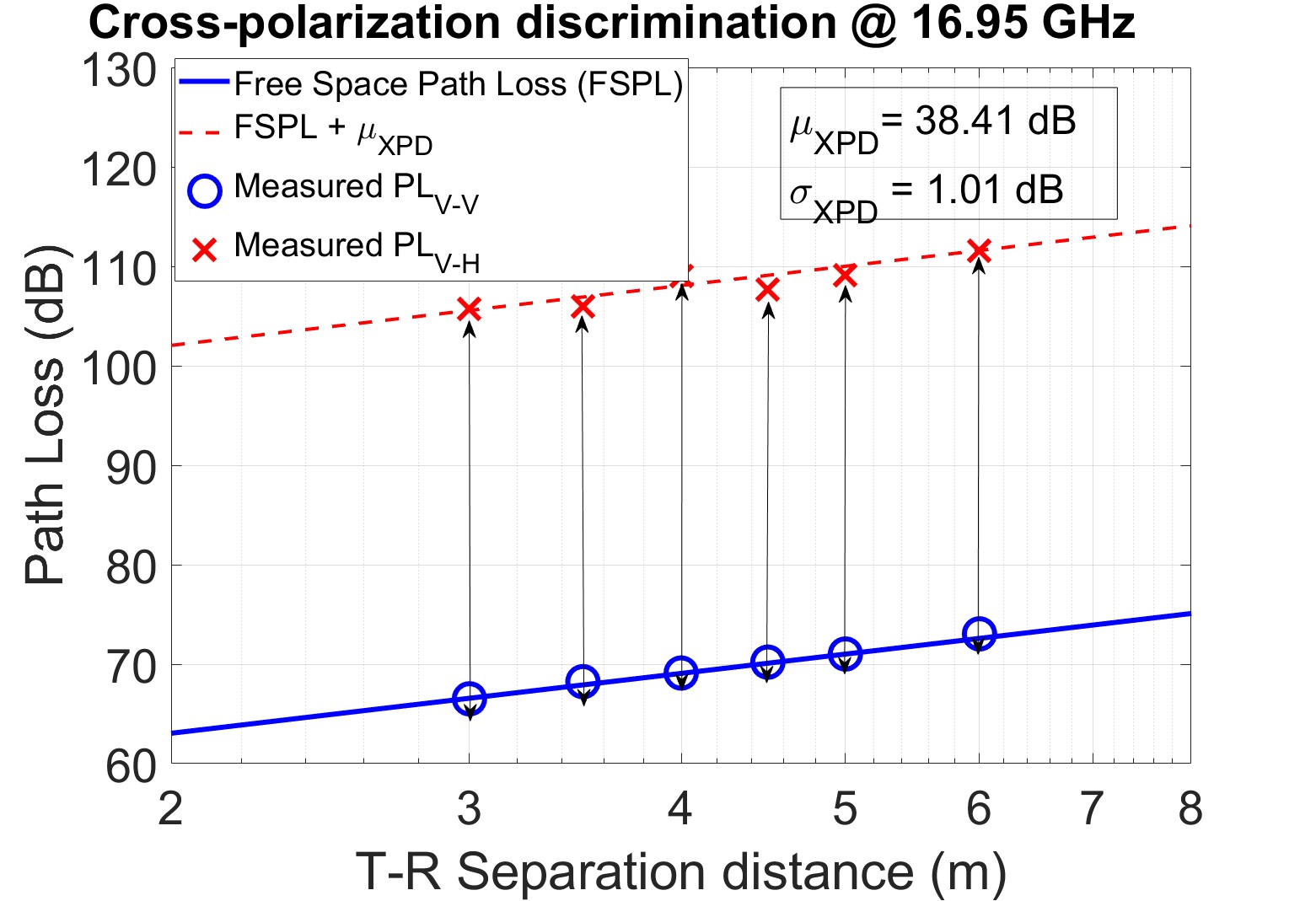}}%
		\\[2.6mm]
		\vspace{-10 pt}
		\caption{Measurement of the antenna XPD at (a) 6.75 GHz with 15 dBi, 30$^{\circ}$ HPBW WR137 horn antennas, and (b) 16.95 GHz with 20 dBi, 15$^{\circ}$ HPBW WR62 horn antennas, using method in \cite{Xing2018vtc}.\newline}
		\vspace{-10 pt}
		\label{fig:XPD}
	\end{figure}

	During propagation, interaction with the environment can cause the transmitted signal energy to be captured in cross-polarized antenna orientation. Particularly, different materials can exhibit varying penetration loss for co and cross-polarized antenna configurations\cite{Xing2018vtc}. Thus, observing the cross-polarization discrimination (XPD) facilitates the characterization of energy captured in the orthogonal orientation. In this study, we used all four possible polarizations during the penetration loss measurements of each material. To accurately determine material pentration loss, the free space (XPD) must be measured. Fig. \ref{fig:XPD} shows the antenna XPD measured for the channel sounder system at (a) 6.75 GHz and (b) 16.95 GHz in V-V and V-H polarizations, using the calibration method in \cite{Xing2018vtc}. Waveguide twists at WR-137 and WR-62 facilitate the change in polarization.

	To obtain empirical XPD for the horn antennas used in this penetration loss campaign, measurements of FSPL are made for the different antenna polarization configurations starting in the far field from T-R separation distances of 3 m with increments of 0.5 m up to 6 m. An open lab area ensured the ground and ceiling multipath were avoided and only LOS path was captured\cite{Newhall1996rd}. The difference in path loss (PL) between the co-polarized (V-V or H-H) and cross-polarized (V-H or H-V) configurations at a T-R separation distance, $d$, yields the XPD, as shown in \eqref{eq:xpd}. A mean XPD of 35.75 dB is observed at 6.75 GHz with $\sigma_{XPD}=1.12 $ dB. Likewise, mean XPD at 16.95 GHz is obtained as 38.41 dB with $\sigma_{XPD}=1.01 $ dB.
	\begin{equation}
		\label{eq:xpd}
		\begin{split}
			XPD(d)[dB]= PL_{V-V}(d)[dB] - PL_{V-H}(d)[dB], \\ 
		\end{split}
	\end{equation}
	%where,
	\begin{equation*}
		%\text{where,}\\ 
		PL [dB] = P_t [dBm] - P_r [dBm] + 
		G_{TX} [dBi] + G_{RX} [dBi]
	\end{equation*} 
	
	$P_t$ in \eqref{eq:xpd} denotes the true transmit power recorded with the power meter during calibration. $P_r$ represents the power in the first arriving multipath in the PDP. $G_{TX}$ and $G_{RX}$ are the antenna gains.

	\section{Measurement procedure for Penetration Loss}
	\label{sxn:msmt}
	The penetration loss, $L$, for a co-polarized or cross-polarized antenna configuration, is determined by observing the difference between the received powers, $P_r$, at identical T-R separation distances, first in free-space and then with the material under test (MUT) in the line-of-sight path of the TX and RX, with a path that is perpendicular to the span of the MUT (e.g. on boresight, or on normal incidence), \eqref{eq:pen}. $P_r$ in \eqref{eq:pen} is obtained as the power received in the first arriving multipath in the captured PDP ignoring the other multipath \cite{newhall1997amtams,durgin:1998:measurements-and-models-for-radio}. The 1 GHz bandwidth channel sounder allows temporal resolution of multipath up to 1 ns apart (e.g 1 foot or ~0.3 m) to ensure that reflected multipath are not within the first arriving LOS multipath component in a PDP and thus avoiding frequency selectivity of narrowband systems \cite{Newhall1996rd}.   
	\begin{equation}
		\label{eq:pen}
		\begin{split}
			L[dB]= P_{r}^{FS}[dB] - P_{r}^{MUT}[dB], \\ 
		\end{split}
	\end{equation}
	
	Measurements are made for three T-R separation distances of 2.0, 2.5, and 3.0 m (all measured distances are in the far field). Following the measurement of $P_{r}^{FS}$, the MUT (the MUT dimensions exceed $15\times\lambda$ in all directions for each MUT) is placed halfway between the TX and RX to ensure the entire wavefront illuminates the MUT. 
	When measuring in-situ partitions, such as a drywall partition in an indoor corridor or wooden doors in rooms that prevented the TX and RX from being spaced equally, the TX is placed at a fixed separation of 1 m, while the RX is moved further at distances ranging from 1 to 2 m from the MUT to achieve the overall T-R separation distance of 2 to 3 m. %A minimum separation of 1 m from the MUT was always maintained for all in-situ partitions to ensure the impinging wavefront was in the far field. 
	Though a separation much greater than 5$\times D_f$  is recommended \cite{Xing2018vtc} for the TX/RX and MUT, the larger antenna dimension and the HPBW at 6.75 and 16.95 GHz makes achieving such separation impractical considering the MUT dimensions, ceiling and antenna height, and available space. 
	
	After positioning the TX and RX on either side of the MUT with a fixed separation, the TX and RX antenna orientations in both co and cross polarization are adjusted to establish a boresight configuration with the strongest first-arriving (LOS) multipath component. To mitigate potential RX antenna misalignment effects due to large HPBWs, the RX antenna is subsequently moved in azimuth and elevation by 1° increments around its boresight, resulting in a total of five PDPs recorded at each RX pointing direction. The penetration loss, $L$, at each distance is obtained by subtracting the linear average of these five recorded powers expressed in dB, $P_{r}^{MUT}$ from $P_{r}^{FS}$, as shown in \eqref{eq:pen}. Once $L$ at each distance is calculated, the average penetration loss for the MUT is recorded as the mean of $L$ in linear scale at the different distances. Thereafter, $P_{r}^{MUT}$ and $P_{r}^{FS}$ are obtained for V-H, H-V, and H-H antenna polarization. Finally, $L_{VV}$ and $L_{HH}$ are linearly averaged to obtain $L$ for co-polarized antennas, and $L_{VH}$ and $L_{HV}$ for cross-polarized antennas. 
	
	\par
\begin{figure}[htbp]
		\centering
		\includegraphics[width=0.4\textwidth]{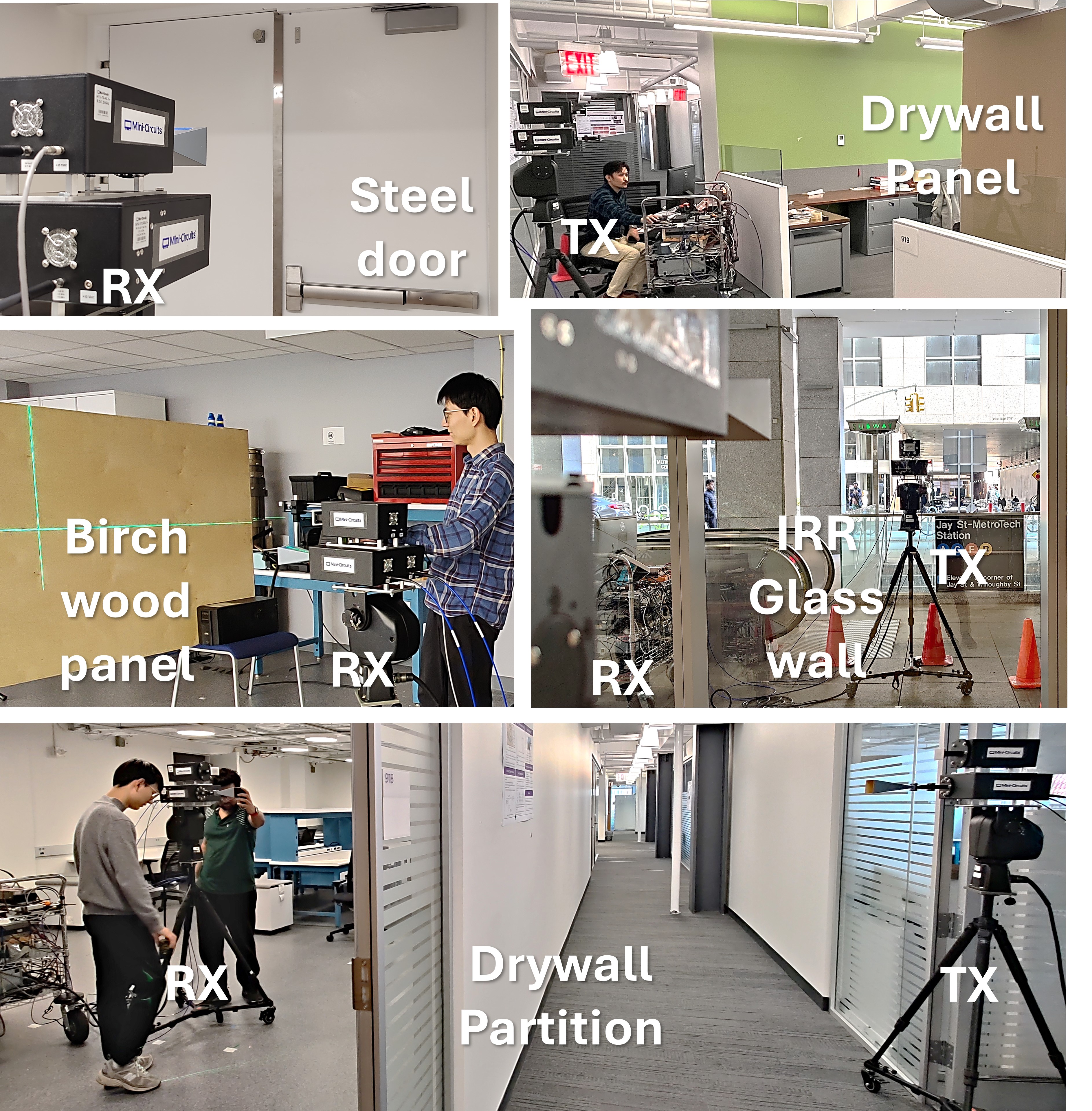}
		\caption{Various MUT measured; Large panels of different materials, such as drywall were measured in a lab setup. Partitions, such as IRR glass walls required TX and RX to be set up on both sides.\newline}
		\label{fig:MUT_msmt}
		\vspace{-30 pt}
	\end{figure}

 Ten commonly found indoor materials are measured. Panel sheets of the all the tested materials are employed to ensure a uniform material upon which the incident wavefront impinges. Images of the materials and partitions being measured are shown in Fig. \ref{fig:MUT_msmt} and specifications are listed below:
	\begin{itemize}
		\item Drywall panel: Penetration through a USG Sheetrock brand gypsum drywall panel is measured. Dimensions of the drywall sheet are 4 ft$\times$8 ft with 3 cm thickness.
		\item Birch Wood panel: A thick Birch plywood panel with 13 plies pressed together is measured. The plywood panel has dimensions of 4 ft$\times$8 ft with 2 cm thickness. 
		\item Whiteboard: A rollable glass whiteboard with front and back writing faces is used. The whiteboard has a tempered glass finish over the laminated backing. Dimensions of the whiteboard are 72"$\times$40" with 3 cm thickness.
		\item Low-emissivity (low-e) or IRR glass window: The window used for the measurement is a 59"$\times$59" sliding panel with low-e double pane glass. Argon gas is filled in the cavity between the two panes and the window has a U-value rating of 0.29 (insulating capability of the glass, lower values indicate greater insulation\cite{Shakya2022icc}). Total thickness of the two panes is 2 cm. The outer frame of the window is white vinyl plastic. 
	\end{itemize}
	
	Other common building partitions and walls such as plasterboard wall, cinderblock wall, and doors found in-situ within the building were also measured. An O2I penetration was measured with the massive glass curtain wall on the perimeter of the 370 Jay Street building. The details of these indoor and O2I partitions are as follows:
	\begin{itemize}
		\item Wooden Door: Large capacity lecture halls at the NYU Tandon School of Engineering have a large wooden double door. The door is made of a fire safety rated solid wood core with 4.5 cm thickness. The door has door handles on one side and a push bar on the other side. The TX and RX heights were raised for the incident wave to impinge on a uniform door surface.  
		\item Steel Door: Hallway entrances and some labs in the 370 Jay Street building at NYU Tandon use large doors with a skin of steel material. The steel door also has a metal push bar at the middle of the door. The TX and RX heights were raised to ensure penetration loss of a uniform steel door (without the metal push bar) was captured. 
		\item Clear glass wall: The measured partition is a single panel transparent glass wall without any coating having 0.5 cm thickness used on the periphery of the Wireless RF Propagation Lab at 370 Jay Street.
		\item Low-e or IRR glass wall: The 370 Jay Street building has a glass curtain wall on the ground floor. Double-pane tinted glass with $1/2$" thick panels and Argon gas filling between the glass panes is used for the curtain wall with individual panels supported by a metal frame. The glass achieves a U-value of 0.26. The total thickness of the glass panels is 3 cm. %Additionally, the glass is tinted to reduce visibility from the outside and enhance thermal efficiency. The glass achieves a U-value of 0.26. The total thickness of the glass panels is 3 cm. 
		\item Cinderblock wall: Few classroom partitions on the NYU campus are cinderblock walls with a coat of paint. Some foam soundboards are placed with irregular spacing on the concrete walls for sound absorption in the classroom. The cinderblock wall measures 22 cm thick.  
		\item Plasterboard wall: The room partitions in the NYU WIRELESS Research Center are composed of sheets of drywall placed on a metal frame. Some plasterboard partitions are also known to have metal studs between the sheets. A stud detector helped ensure no studs were present in the region of the measurement. 
	\end{itemize}

	%The standard deviation of the penetration loss, measured over distance for both co-polarized and cross-polarized antennas, is less than 2.64 dB. The average penetration loss in both cases is approximately xxxx dB (xxxx dB/cm).
	
	\section{Penetration Measurement Results}
	\label{sxn:res}
	The penetration loss measurement results are presented in Table \ref{tab:penLoss}. For both co-polarized and cross-polarized antennas, the measured penetration loss exhibits minimal variation across the employed T-R separation distances. Co-pol penetration loss is computed by linearly averaging $L$ evaluated for V-V and H-H, while cross-pol penetration loss is obtained as the linear average of $L$ for V-H and H-V, as detailed in Section \ref{sxn:msmt}.
	% The co-pol penetration loss is calculated by averaging the measurement results from V-V and H-H polarizations. Similarly, the cross-pol penetration loss is determined by averaging the results from V-H and H-V polarizations, as explained in Section \ref{sxn:msmt}.
	% The penetration loss for each polarization is obtained by linearly averaging between $L$ calculated for V-V and V-H, as explained in Section \ref{sxn:msmt}. 
	The penetration loss at 6.75 GHz is consistently lower compared to 16.75 GHz for both co and cross-polarized configurations. The largest loss measured is for the steel door exhibiting over 40 dB and 50 dB penetration loss at the 6.75 GHz and 16.95 GHz frequencies, respectively. The low-e glass curtain wall, used on the building outer surface, also exhibits strong attenuation of over 30 dB at both measurement frequencies for co and cross polarizations. The low-e window also shows significant attenuation, interestingly, loss for the cross-polarized configuration is much lower compared to the co-polarized measurements (29.7 dB Co Pol; 15.3 dB Cross Pol @ 6.75 GHz, and 32.68 dB Co Pol; 19.52 dB Cross Pol @ 16.95 GHz). This indicates a strong dependence on polarization for low-e window glass attenuation.     
	
	\renewcommand{\arraystretch}{1.2}
	\begin{table}[!t]
		\centering
		\caption{Measured Penetration Loss of materials for Co-pol and Cross-pol for FR1(C) and FR3}
		\begin{tabular}{|c|@{\hspace{1 pt}}p{0.6 cm}|c@{\hspace{1 pt}}|p{0.4 cm}|p{0.4 cm}?p{0.4 cm}|p{0.4 cm}?p{1.1 cm}|}
			\hline
			\multicolumn{1}{|c|}{\multirow{3}{1.5 cm}{\centering \textbf{MUT}}} & \multicolumn{1}{@{\hspace{1 pt}}p{0.6 cm}|}{\multirow{3}{0.6 cm}{\textbf{Width (cm)}}} & \multirow{3}{*}{\textbf{Pol}} & \multicolumn{4}{c?}{\textbf{Frequency}} & \\
			\cline{4-7}    \multicolumn{1}{|c|}{}      &      & \multicolumn{1}{c|}{} & \multicolumn{2}{c?}{\textbf{6.75 GHz}} & \multicolumn{2}{c?}{\textbf{16.95 GHz}}& \textbf{$\Delta_{\mu}$ (dB)}\\
			\cline{4-7}    \multicolumn{1}{|c|}{}     &       & \multicolumn{1}{c|}{} & \multicolumn{1}{p{0.4 cm}|}{\textbf{$\mu$ (dB)}} & \multicolumn{1}{p{0.4 cm}?}{\textbf{$\sigma$ (dB)}} & \multicolumn{1}{p{0.4 cm}|}{\textbf{$\mu$ (dB)}} & \multicolumn{1}{p{0.4 cm}?}{\textbf{$\sigma$ (dB)}}& $_{16.95-6.75}$ \scriptsize{GHz}\\
			\specialrule{1pt}{0pt}{0pt}
			\multicolumn{1}{|c|}{\multirow{2}{1.5 cm}{\centering Cinderblock Wall}} & \multirow{2}{0.6 cm}{\centering 22} &Co   & 13.4  & 1.6  & 15.0 & 1.1& \centering\arraybackslash1.6 \\
			\cline{3-8}    \multicolumn{1}{|c|}{}      &       & Cross   & 10.7 & 1.6  & 11.5 & 0.7& \centering\arraybackslash0.9 \\
			\specialrule{1pt}{0pt}{0pt}
			\multicolumn{1}{|c|}{\multirow{2}{1.5 cm}{\centering Low-e tinted glass wall}} & \multirow{2}{0.6 cm}{\centering 3} & Co   & 33.7 & 0.9  & 42.3 & 0.2& \centering\arraybackslash8.6\\
			\cline{3-8}    \multicolumn{1}{|c|}{}      &       & Cross   & 38.4 & 1.2  & 46.5 & 1.0& \centering\arraybackslash8.2\\
			\specialrule{1pt}{0pt}{0pt}
			\multicolumn{1}{|c|}{\multirow{2}{1.5 cm}{\centering Low-e glass window}} & \multirow{2}{0.6 cm}{\centering 2} & Co   & 29.7 & 1.4   & 32.7 & 2.6& \centering\arraybackslash3.0\\
			\cline{3-8}    \multicolumn{1}{|c|}{}      &       & Cross   & 15.4  & 0.7  & 18.5 & 2.3& \centering\arraybackslash3.1\\
			\specialrule{1pt}{0pt}{0pt}
			\multicolumn{1}{|c|}{\multirow{2}{1.5 cm}{\centering Clear Glass}} & \multirow{2}{0.6 cm}{\centering 1} & Co   & 3.6  & 0.3  & 3.7  & 0.4& \centering\arraybackslash0.1\\
			\cline{3-8}     \multicolumn{1}{|c|}{}     &       & Cross   & 4.2  & 0.4  & 4.4 & 0.6& \centering\arraybackslash0.2\\
			\specialrule{1pt}{0pt}{0pt}
			\multicolumn{1}{|c|}{\multirow{2}{1.5 cm}{\centering Birch Wood panel}} & \multirow{2}{0.6 cm}{\centering 2} & Co   & 2.4  & 0.7  & 6.1  & 0.4& \centering\arraybackslash3.7\\
			\cline{3-8}    \multicolumn{1}{|c|}{}      &       & Cross   & 2.0  & 0.9  & 5.5  & 0.9& \centering\arraybackslash3.5\\
			\specialrule{1pt}{0pt}{0pt}
			\multicolumn{1}{|c|}{\multirow{2}{1.5 cm}{\centering Wooden door}} & \multirow{2}{0.6 cm}{\centering 4.5} & Co   & 5.8  & 1.1  & 6.1  & 1.3& \centering\arraybackslash0.3\\
			\cline{3-8}    \multicolumn{1}{|c|}{}      &       & Cross   & 7.1 & 2.2  & 7.7  & 0.8& \centering\arraybackslash0.6\\
			\specialrule{1pt}{0pt}{0pt}
			\multicolumn{1}{|c|}{\multirow{2}{1.5 cm}{\centering Steel door}} & \multirow{2}{0.6 cm}{\centering 4.7} & Co   & 43.2  & 0.5  & 58.5 & 1.4& \centering\arraybackslash15.3\\
			\cline{3-8}    \multicolumn{1}{|c|}{}      &       & Cross  & 41.8 & 0.8  & 56.4 & 1.7& \centering\arraybackslash14.6\\
			\specialrule{1pt}{0pt}{0pt}
			\multicolumn{1}{|c|}{\multirow{2}{1.5 cm}{\centering Plasterboard wall}} & \multirow{2}{0.6 cm}{\centering 13.7} & Co   & 2.1  & 1.0  & 4.5  & 0.4& \centering\arraybackslash2.4\\
			\cline{3-8}    \multicolumn{1}{|c|}{}      &       & Cross   & 3.0  & 0.6  & 6.1  & 1.2& \centering\arraybackslash3.2\\
			\specialrule{1pt}{0pt}{0pt}
			\multicolumn{1}{|c|}{\multirow{2}{1.5 cm}{\centering Drywall panel}} & \multirow{2}{0.6 cm}{\centering 3} & Co   & 0.6  & 0.1   & 1.2  & 0.5& \centering\arraybackslash0.6\\
			\cline{3-8}    \multicolumn{1}{|c|}{}      &       & Cross   & 1.5  & 0.2  & 2.3  & 1.2& \centering\arraybackslash0.8\\
			\specialrule{1pt}{0pt}{0pt}
			\multicolumn{1}{|c|}{\multirow{2}{1.5 cm}{\centering White Board}} & \multirow{2}{0.6 cm}{\centering 3} & Co   & 3.1  & 0.3  & 6.9  & 0.5& \centering\arraybackslash3.8\\
			\cline{3-8}    \multicolumn{1}{|c|}{}      &       & Cross   & 4.1 & 0.7  & 7.5  & 1.3& \centering\arraybackslash3.4\\
			\specialrule{1pt}{0pt}{0pt}
		\end{tabular}%
		\label{tab:penLoss}%
		\vspace*{-1\baselineskip}
		\vspace{-10 pt}
	\end{table}
	\renewcommand{\arraystretch}{1.0}
	
\begin{figure}[!t]
		\centering
		% First subfigure equivalent
		\begin{minipage}{0.48\textwidth}
			\centering
			\includegraphics[width=0.94\textwidth]{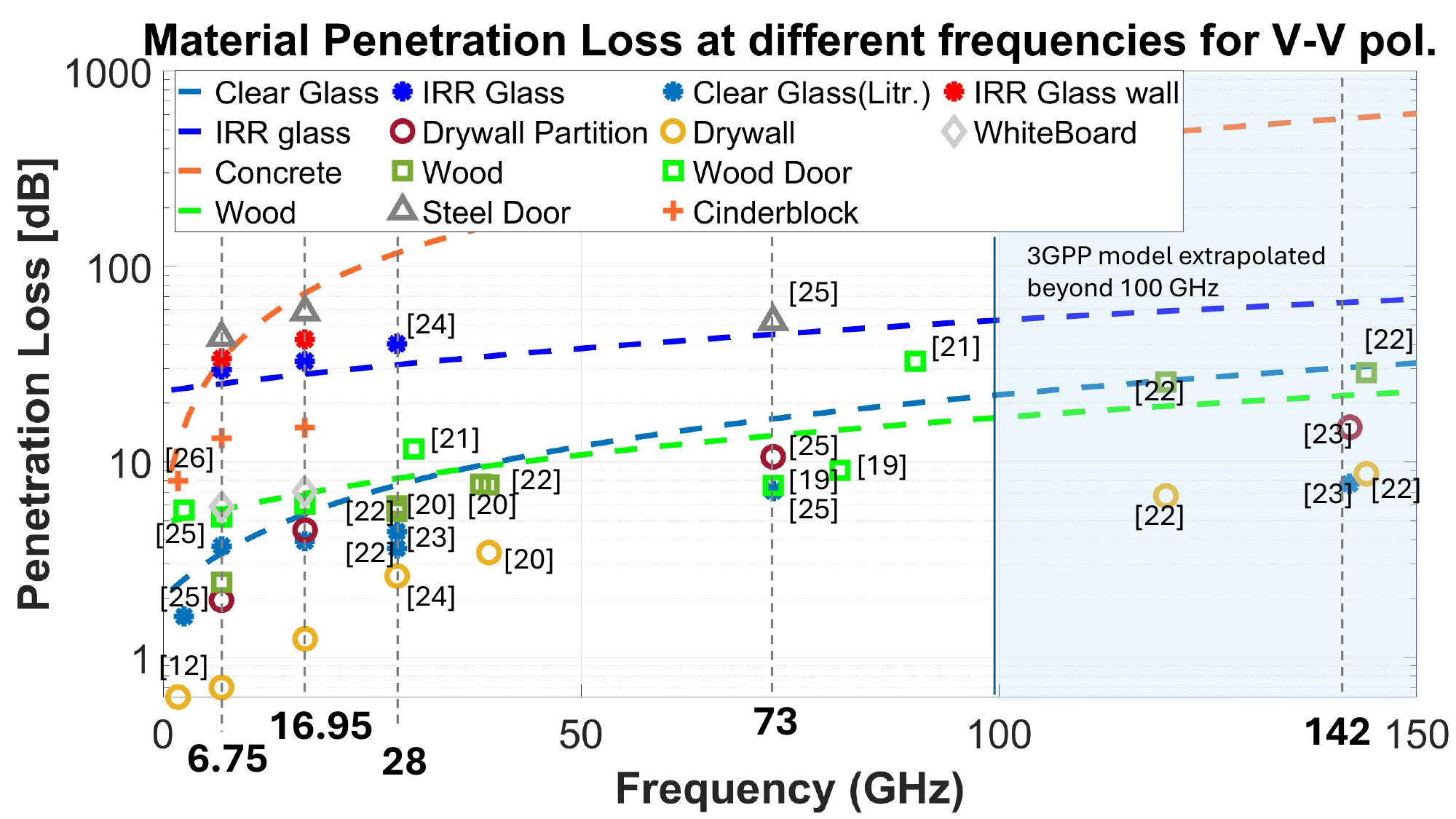}
			% Manual caption
			(a)
			%(a) Penetration loss of various materials at FR3, mmWave, and sub-THz frequencies and comparisons with the 3GPP model.
			\label{fig:MUTs-a}
		\end{minipage}
		\hspace{0.04\textwidth} % optional: space between the images
		% Second subfigure equivalent
		\begin{minipage}{0.46\textwidth}
			\centering
			\includegraphics[width=0.88\textwidth]{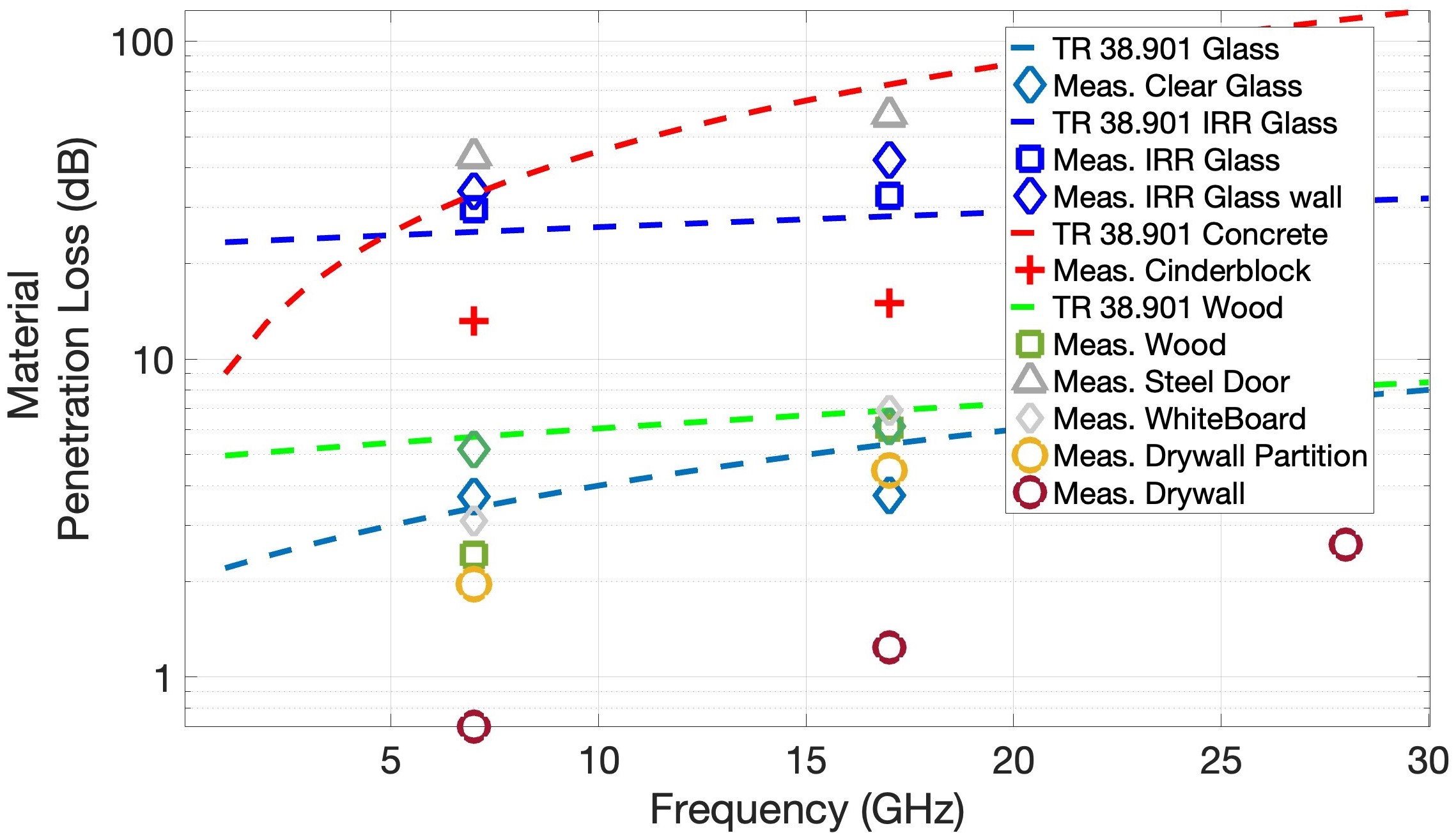}
			% Manual caption
			\\(b)
			%(b) Zoomed view of the penetration loss in the 0-30 GHz range.
			\label{fig:MUTs-b}
		\end{minipage}
		\caption{Penetration loss of various materials at different frequencies compared with the 3GPP TR-38.901 material penetration models. (a) Range 0-150 GHz. (b) Range 0-30 GHz.}
		\label{fig:MUTs}
		\vspace{-15 pt}
	\end{figure}

	% Fig. \ref{fig:MUTs} shows the penetration loss of different materials, collected from past measurements made at NYU up to 142 GHz, along with values reported in the literature at sub-6 GHz, mmWave and sub-THz frequencies\cite{Khatun2019gc,Vargas2018apwc,Matolak2021wamicon,Du2021icc,Xing2018gc,Zhao2013icc,Ryan2017icc,Du2016gcw,Rappaport1994apm,zhang:1994:measurements-of-the-characteristics-of}. The dB/cm normalized attenuation is used from the values reported in the literature and multiplied with the thickness of materials used here to plot points in Fig. \ref{fig:MUTs}. 	
	Fig. \ref{fig:MUTs} (a) depicts the penetration loss of ten materials investigated in this paper at specific FR1C and FR3 frequencies, along with their penetration loss values in sub-6 GHz, mmWave, and sub-THz bands, as reported in  \cite{Khatun2019gc,Vargas2018apwc,Matolak2021wamicon,Du2021icc,Xing2018gc,Zhao2013icc,Ryan2017icc,Du2016gcw,Rappaport1994apm,zhang:1994:measurements-of-the-characteristics-of,Shuai2013pimrc}. To facilitate direct comparison of the material penetration loss across the entire frequency range of 0.5-150 GHz, the dB/cm penetration loss values for each material is multiplied by their respective thicknesses. The comparisons reveal an increasing penetration loss with increasing frequency for all ten of the materials measured. For example, the yellow rings for drywall in Fig. \ref{fig:MUTs}(a) shows a clear increase with frequency when measurements from this paper are augmented with values reported in \cite{Matolak2021wamicon,Khatun2019gc,Xing2018gc}. 
	% Notably, the measured penetration loss for materials such as clear glass and IRR glass closely follows the model's trend, albeit with some variability. In contrast, materials like cinderblock and drywall partition exhibit higher variability from the trend, suggesting a frequency-dependent loss characteristic that diverges from the model at specific points within the analyzed range. Such insights into the behavior of materials at these frequencies are essential for developing accurate models for wireless communication systems, particularly in urban environments where these materials are ubiquitous.
	On the other hand, Fig. \ref{fig:MUTs} (b) only shows the penetration loss from the measurements conducted in this paper at 6.75 and 16.95 GHz.

	\section{Comparison with 3GPP Material Penetration Loss Model}
	\label{sxn:Comp}
	Penetration loss data in the 7--24 GHz range is sparse in the literature. Recent 3GPP discussions emphasize a critical gap in channel measurement and O2I loss data for the 7--24 GHz band. Over 80\% of the channel measurement data submitted in 3GPP falls outside the 7--24 GHz band, concentrated below 6 GHz or exceeding 28 GHz \cite{rp234018}. Thus, validation of the 3GPP TR 38.901 model for material penetration loss within the 7--24 GHz range is needed.
	
	\par The current 3GPP model for material penetration loss (Table 7.4.3-1, 3GPP TR 38.901 Release 18 \cite{3GPPTR38901}) provides model paramters for concrete walls, IRR glass, standard multipane glass (clear glass), and wood that vary linearly with frequency. The Root Mean Square Error (RMSE) calculated between the measured and the 3GPP predicted penetration losses for the materials at 6.75 GHz and 16.95 GHz, demonstrates a close adherence to the 3GPP model for conventional materials like clear glass (RMSE = 1.2 dB) and wood (RMSE = 2.4 dB). This indicates the 3GPP model validity for wood and standard multi-pane glass. Conversely, concrete walls and IRR glass walls exhibit significantly higher RMSE values (42.9 dB and 11.8 dB, respectively) at both frequencies. The 3GPP model consistently under-predicts the loss for IRR glass. However, for concrete walls, the observed discrepancy may be attributed to measurements in this paper characterizing penetration through an indoor cinderblock wall, which differs substantially from the thicker building exterior walls considered by the 3GPP model. Results suggest revisions may be required for 3GPP material penetration models for IRR glass and concrete \cite{r12402407}.

	\section*{Conclusion}
	The empirical propagation penetration loss exhibited by common materials found indoors and on building perimeters was reported in the paper. The measurements were made with a wideband sliding correlation-based channel sounder operating at 6.75 GHz and 16.95 GHz. %The antenna XPD was characterized for the horn antennas at both the frequency bands and the mean was obtained to be 35.75 dB at 6.75 GHz and 38.41 dB at 16.95 GHz. 
	Penetration loss was consistently observed to be lower at 6.75 GHz compared to 16.75 GHz for co and cross-polarized antennas encompassing V-V, V-H, H-V, and H-H configurations. Low-e glass windows exhibited polarization-dependent attenuation of the signal. Upon comparing with penetration loss measurements at mmWave and sub-THz frequencies, the materials exhibited increasing loss at higher frequencies. Comparisons with the 3GPP models for material penetration show close adherence for wood and clear glass, while IRR glass and concrete showed RMSE errors above 10 dB, suggesting revisions may be required to current models. Reported material characteristics provide valuable information for future 5G and 6G wireless systems in the FR1(C) and FR3 upper mid-band spectrum. 
	
	\section*{Acknowledgment}
	Authors thank Profs. Sundeep Rangan and Marco Mezzavilla for support with experimental licensing in the FR3 bands. %Authors acknowledge the NYU staff and colleagues for their support during measurements around campus.
	
	\bibliographystyle{IEEEtran}
	\bibliography{references}
	
\end{document}